\documentclass[letterpaper, 10 pt, conference]{ieeeconf}     

\IEEEoverridecommandlockouts                              

\usepackage{authblk}
\usepackage{amsfonts}
\usepackage{amsmath}
\usepackage{hyperref,cite}
\usepackage{cleveref}
\usepackage{color}
\usepackage{algorithm}
\usepackage{algorithmic}
\let\labelindent\relax
\usepackage{enumitem}
\usepackage{float}
\usepackage{graphicx}
\usepackage{booktabs}
\usepackage{threeparttable}
\usepackage{subfigure}

\begin{document}
\title{\LARGE \bf
A rapid-prototype MPC tool based on gPROMS platform}
\author[1,2]{Liang Wu}
\author[2]{Maarten Nauta}

\affil[1]{IMT School for Advanced Studies Lucca, Italy}
\affil[2]{Siemens Process Systems Enterprise Ltd., London (UK)}
\thispagestyle{empty}
\pagestyle{empty}
\maketitle

\begin{abstract}
This paper presents a rapid-prototype Model Predictive Control (MPC) tool based on the gPROMS platform, with the support for the whole MPC design workflow. The gPROMS-MPC tool can not only directly interact with a first-principle-based gPROMS model for closed-loop simulations but also utilizes its mathematical information to derive simplified control-oriented models, basically via linearization techniques. It can inherit the interpretability of the first-principle-based gPROMS model, unlike the PAROC framework in which the control-oriented models are obtained from black-box system identification based on gPROMS simulation data. The gPROMS-MPC tool allows users to choose when to linearize such as at each sampling time (successive linearization) or some specific points to obtain one or multiple good linear models. The gPROMS-MPC tool implements our previous construction-free \textit{CDAL} and the online parametric active-set \textit{qpOASES} algorithms to solve sparse or condensed MPC problem formulations, respectively, for possible successive linearization or high state-dimension cases. Our \textit{CDAL} algorithm is also matrix-free and library-free, thus supporting embedded C-code generation. After many example validations of the tool, here we only show one example to investigate the performance of different MPC schemes.
\end{abstract}


\section{Introduction}\label{sec:intro}
Developing a novel chemical or pharmaceutical process can take decades and hundreds of millions of dollars, and the first-principle-based modeling or digital-twin technology can speed up development and reduce costs. The gPROMS' digital-twin technology (http://www.psenterprise.com/gproms) \cite{gPROMS2022} provides powerful and easy-to-use process modeling capabilities, rich physical property databases and model libraries, integrated experimental design, parameter estimation, and optimization capabilities. 

\begin{figure}
\centering
        \hspace*{-1em}\includegraphics[width=1\columnwidth]{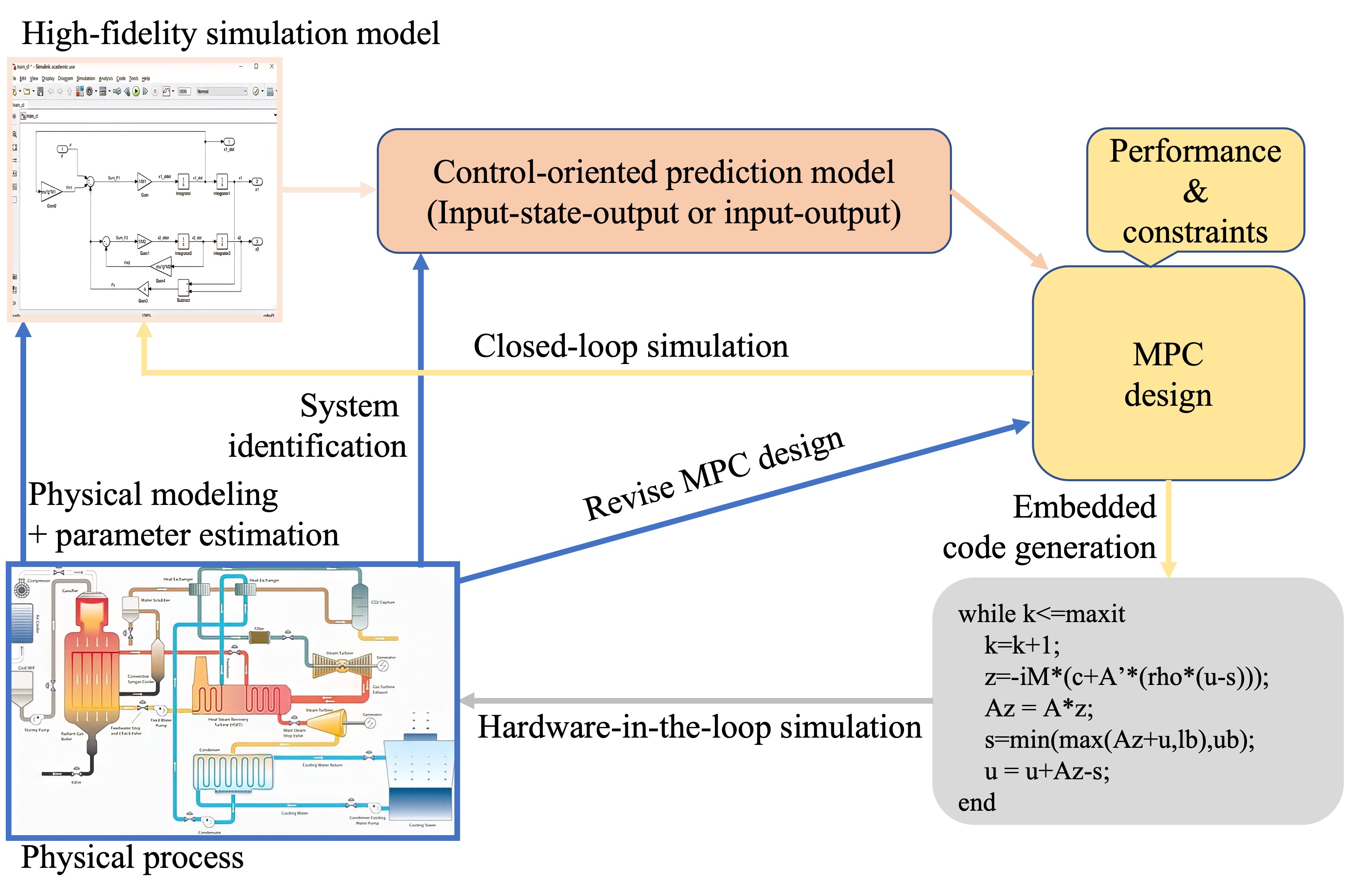} 
        \caption{The schematic diagram of whole MPC design flow}
        \label{MPC_design_flow} 
\end{figure}

During the process design phase, the process sometimes requires a controller to maintain stability or simultaneously is explored together with the controller design.
The process often appears as a multi-input multi-output (MIMO) plant, and using a multi-PID controller scheme would be cumbersome or complicated for relatively inexperienced process engineers. Model predictive control (MPC) is developed for controlling MIMO plants subject to constraints \cite{morari1999model}, and MPC has been widely used in diverse industrial areas, such as process \cite{qin2003survey}, and aerospace \cite{eren2017model}, power electronics \cite{geyer2016model}, etc. 
The design workflow of an MPC controller for a physical plant is shown in Fig. \ref{MPC_design_flow}. Firstly, with the help of professional process modeling software such as gRPOMS, Aspen Plus \cite{Aspen2022}, or general-purpose modeling software MATLAB/Simulink \cite{matlabSimulink}, a high-fidelity model based on first principle is established. The first-principle-based model could be as complex as possible, like introducing distributed parameter model to describe spatial-time relationships, which is usually a large mixed system of integral, partial differential, and algebraic equations (IPDAEs) \cite{oh1996modelling}. The first-principle-based model can be calibrated from the experimental data with the use of parameter estimation or validated. The main objective of developing a first-principle-based model is to explore the optimal design space of the physical process, and another objective is to validate the designed controller scheme via closed-loop simulations. A key part of MPC design is how to obtain a simplified control-oriented prediction model of the physical process to predict its likely evolution. One approach is to use system identification method to obtain a simplified model from experimental data, such as the open-loop step-response data. Getting open-loop experimental data is sometimes expensive or even forbidden for safety reasons. As an alternative, the simulation data of first-principle-based models can be used for system identification. Another approach is based on the linearization of first-principle-based model at some operating points. 

\subsection{Related works}\label{sec:related_works}
In the PAROC framework \cite{pistikopoulos2015paroc} developed by Pistikopoulos et.al, simulation data are generated from first-principle-based gPROMS models and then used by MATLAB's system identification and model reduction toolbox to obtain multiple approximate linear state-space models. Based on these linear state-space models, explicit/multi-parametric MPC algorithm \cite{bemporad2002model, bemporad2002explicit} are implemented on MATLAB side. The closed-loop simulations are performed via the the gPROMS ModelBuilder\textsuperscript{\textregistered} tool gO:MATLAB, which couples the computation of the gPROMS and MATLAB. Although the PAROC framework takes advantage of MATLAB's power in MPC design and 
gPROMS' benefits in professional process modeling, the PAROC framework obviously does not take full advantage of the mathematical information of existing first-principle-based gPROMS models. The control-oriented model obtained by their method is still a black-box model and does not inherit the interpretability of the first-principle-based gPROMS model. By directly linearizing the first-principle-based gPROMS model, we can obtain simplified control-oriented models with interpretability.
In the previous \textit{gNLMPC} \cite{pfeiffer2020nonlinear}, the nonlinear gPROMS model is directly used to construct a nonlinear MPC problem, which is solved by using the built-in numerical optimization capabilities of the gPROMS platform. Clearly, the \textit{gNLMPC} would have good closed-loop performance but at a high online computation cost. In fact, 
linear model-based MPC are successful in the vast majority of process industry applications even though many manufacturing processes are inherently nonlinear \cite{qin2000overview}. 

\subsection{Contribution}\label{sec:contribution}
This paper presents the development of a rapid-prototype MPC tool based on the gPROMS platform. The gPROMS-MPC tool employs the online successive linearization strategy to linearize a general nonlinear first-principle-based gPROMS model, based on the supporting automatic differentiation capability of gPROMS model. The gPROMS-MPC tool implements our previous construction-free \textit{CDAL} algorithm \cite{wu2021simple} and the widely-used open-source \textit{qpOASES} v3.2 algorithm \cite{ferreau2014qpoases} in the gPROMS platform (gPROMS Process v2.2.2), based on sparse and condensed MPC formulations, respectively. 

The online successive linearization strategy is an effective solution to deal with nonlinear MPC problems \cite{camacho2007nonlinear,falcone2007predictive}, by achieving a good trade-off between computational cost and closed-loop control performance. In addition to linearization at each sampling time, the gPROMS-MPC tool allows users to decide when to perform the linearization at a lower sampling frequency or at some specific points to derive one or multiple invariant linear state-space models in embedded MPC design. The embedded MPC design is the final step of the MPC design workflow shown in Fig. \ref{MPC_design_flow}, that is, evaluating whether the chosen one or multiple invariant linear state-space models meets the closed-loop performance requirements under the computation limits from embedded platforms. Since our construction-free \textit{CDAL} algorithm is also matrix-free and library-free, which makes the gPROMS-MPC tool suitable for the embedded code generation. The gPROMS-MPC is competent in completing all the workflow shown in Fig. \ref{MPC_design_flow} only on the gPROMS platform.

\subsection{Structure and Notation}
The structure of the paper is as follows. Section \ref{sec:mpc_problem_formulation} introduces the general gPROMS model, its subsection \ref{sec:linearized_ss_with_minimal_subset} presents the continuous-time linearized state-space model with minimal subset and formulates the corresponding MPC tracking problem. Further, its subsection \ref{sec:condensed_sparse_formulation} shows the sparse and condensed QP formulations associated with MPC, which are solved by \textit{CDAL} and \textit{qpOASES} v3.2 algorithm, respectively. The two algorithms are shortly illustrated in Section \ref{sec:implementation_gMPC}. In Section \ref{sec:application_example}, a flash-separation example is presented. Finally, we draw conclusions in Section \ref{sec:conclusion}.

$H \succ 0$ ($H \succeq 0$) denotes positive definiteness (semi-definiteness) of a square matrix $H$. For a vector $z$, $\|z\|_{H}^{2}$ denotes the operation $z^{\prime}Hz$. $H^{\prime}$ (or $z^{\prime}$) denotes the transpose of matrix $H$ (or vector $z$).

\section{MPC problem formulation}\label{sec:mpc_problem_formulation}
A gPROMS model typically comprises mixed sets of non-linear differential and algebraic equations, which can be written in the form
\begin{equation}\label{nonlinearDAE}
    f(\dot x,x,y,u)=0
\end{equation}
where $x(t)$ and $y(t)$ are the sets of differential and algebraic variables respectively (both of which are unknowns to be determined by the gPROMS simulation) while $\dot x(t)$ are the derivatives of $x(t)$ with respect to time $t$, $u$ is the set of input variables that are given functions of time.
Now consider the current time point $(x_c,y_c,u_c)$ on the simulation trajectory, a linear model can be obtained by linearising the Eqn (\ref{nonlinearDAE}) at this point,
\begin{equation}\label{linearized_full_ss}
    \frac{\partial f}{\partial x}\delta x+ \frac{\partial f}{\partial \dot x}\dot{\delta x} + \frac{\partial f}{\partial y} \delta y + \frac{\partial f}{\partial u} \delta u = 0
\end{equation}
where $\delta x = x - x_c$, $\delta y = y - y_c$ and $\delta u = u - u_c$.
\subsection{Linearized State-Space model with minimal subset of differentials}\label{sec:linearized_ss_with_minimal_subset}
Most first-principle-based gPROMS models involve thousands of equations and variables. When performing linearization, by specifying a set of input variables $U$ and output variables $Y$, gPROMS provides the automatic variables-reduction technique according to variables dependencies. Namely, gPROMS generates the following linearized model,
\begin{equation}\label{minimal_linearized_full_ss}
\begin{aligned}
    \dot{\delta X}&= A \delta X + B \delta U 
    \\
    \delta Y &= C \delta X + D \delta U
\end{aligned}
\end{equation}
where the state variables $X$ are determined automatically by gPROMS's built-in \textit{Linearise} as the minimal subset of $x$ that is necessary to express the effects of the specified inputs $U$ on the the specified outputs $Y$ via relationships of the above Eqn (\ref{linearized_full_ss}).

Before applied in MPC problem, the continuous-time state-space model (\ref{minimal_linearized_full_ss}) requires to be discretized to discrete-time form. The gPROMS-MPC tool currently utilizes the one-step Euler discretization method. Given a sampling time $T_s$, we have the following discrete-time state-space model,
\begin{equation}\label{discreteSS}
\begin{aligned}
    \delta X_{t+1} &= A_d \delta X_t + B_d \delta U_t + e\\
    \delta Y_t &= C_d \delta X_t + D_d \delta U_t
\end{aligned}    
\end{equation}
where $A_d=I+T_s A$, $B_d=T_s B$, $C_d = C$, $D_d=D$, $e=T_s f(x_c,\dot x_c, y_c, u_c)$.
The gPROMS-MPC tool currently considers MPC tracking problems, thus a $\Delta$-formulation augmented state-space model is used,
\begin{equation*}
\begin{aligned}
    \hat{X}_{t+1} &= \hat{A} \hat{X}_{t} + \hat{B} \hat{U}_t + \hat{e}\\
    \Delta Y_t &= \hat{C}\hat{X}_{t}
\end{aligned}    
\end{equation*}
where $\hat{X}_t=\left[\begin{array}{c}
    \delta X_t  \\
    \delta U_{t-1} 
\end{array}\right]$, $\hat{U}_t=\Delta\delta U_t$ denotes the increment of $\delta U_t$, $\hat{B}=\left[\begin{array}{c}
    B_d\\
    I 
\end{array}\right]$, $\hat{e}=\left[\begin{array}{c}
    e\\
    0 
\end{array}\right]$, $\hat{C}=\left[\begin{array}{cc}
        C_d & 0
    \end{array}\right]$.

Currently, the gPROMS-MPC tool considers the case that box constraints are subject to the increased input, the input and the output, the MPC tracking problem is shown below,
\begin{eqnarray}\label{MPC}
\min && \frac{1}{2}\sum_{t=0}^{T-1}\left\|\left(\delta Y_{t+1}-\delta r\right)\right\|_{W^{y}}^{2}+\left\| \Delta \hat{U}_{t}\right\|_{W^{\Delta u}}^{2}\nonumber\\
\text {s.t.} && \hat{X}_{t+1}=\hat{A}\hat{X}_{t} + \hat{B}\Delta \hat{U}_t + \hat{e},t=0,1,\ldots,T-1\nonumber\\
&&\delta Y_{t+1} = \hat{C}\hat{X}_{t+1},t=0,\ldots,T-1\nonumber\\
&& \Delta U_{\min } \leq \Delta \hat{U}_{t} \leq \Delta U_{\max }, t=0,\ldots,T-1\nonumber\\
&& U_{\min }-U_c \leq \delta U_{t} \leq U_{\max}-U_c,t=0,\ldots,T-1\nonumber\\
&& Y_{\min }-Y_c \leq \delta Y_{t} \leq Y_{\max }-Y_c,t=1,\ldots,T\nonumber\\
&& \hat{X}_0 = 0
\end{eqnarray}
where $\delta r=r-Y_c$ and $r$ denotes the desired tracking set-points. $W^y \succeq 0$ and $W^{\Delta u} \succeq 0$ denotes the weights matrices of the output and input, respectively. Note that the initial values of $\hat{X}$ are zero thanks to their definition. $\left[\Delta U_{\min},\Delta U_{\max} \right]$, $\left[U_{\min},U_{\max} \right]$ and $\left[Y_{\min},Y_{\max} \right]$ denote the box-constraints of the specified increased input, input and output, respectively.

\subsection{Condensed and sparse formulation}
\label{sec:condensed_sparse_formulation}
Solving the above MPC tracking problem (\ref{MPC}) requires constructing it into a quadratic programming (QP) problem. There are two kinds of MPC-to-QP construction, one is the condensed construction that eliminates the states and another is the sparse construction that keeps the states in the resulting QP problem. Eliminating all states from the MPC problem (\ref{MPC}) to yield a smaller-scale, condensed QP problem of the form:

\begin{eqnarray}
\min && \frac{1}{2}z^{\prime}H_c z+h_c^{\prime}z\nonumber\\
\text {s.t.}  && g_c^l \leq G_c z \leq g_c^u\nonumber\\
&& z_c^l \leq z \leq z_c^u\label{condensedQP}
\end{eqnarray}
where the vector of decision variables only comprises the increase control inputs, $z\stackrel{\text { def }}{=}\left[\Delta \hat{U}_0^{\prime},\Delta \hat{U}_1^{\prime},\cdots, \Delta \hat{U}_{T-1}^{\prime}\right]^{\prime}$, and its Hessian matrix $H_c$ is dense. 

The problem (\ref{MPC}) can also yield a structured QP problem by keeping the discrete-time state-space model as equality constraints,
\begin{eqnarray}
\min && \frac{1}{2}z^{\prime}H_s z+h_s^{\prime}z\nonumber\\
\text {s.t.}  && B_s z = b_s\nonumber\\
&& g_s^l \leq G_s z \leq g_s^u\nonumber\\
&& z_s^l \leq z \leq z_s^u\label{sparseQP}
\end{eqnarray}
where the vector of decision variables comprise both the states and the increased control inputs $z=\stackrel{\text { def }}{=}\left[\Delta \hat{U}_0^{\prime},\hat{X}_1^{\prime},\Delta \hat{U}_1^{\prime},\hat{X}_2^{\prime},\cdots, \Delta \hat{U}_{T-1}^{\prime},\hat{X}_{T}^{\prime}\right]^{\prime}$, and
and its Hessian matrix $H_s$ is sparse. Whether the condensed formulation (\ref{condensedQP}) or the sparse formulation (\ref{sparseQP}) is more computationally advantageous, mainly depends on the number of the states $n_x$, the control input $n_u$ and the length of the prediction horizon $T$ \cite{kouzoupis2015towards}. The condensed formulation (\ref{condensedQP}) is obviously preferable when the number of states is large, which would often be the case when spatial-temporal equations are involved. If the ratio $\frac{n_x}{n_u}$ is small and the prediction horizon is long, the sparse formulation (\ref{sparseQP}) is often more efficient. Moreover, the choice may also depends on whether the condensing procedure (the state elimination) requires to be done once offline (like the LTI MPC case) or performed online, as is the case where the gPROMS-MPC tool utilizes the online successive linearization strategy for nonlinear systems. Thus, the gPROMS-MPC tool implements the two formulations to allow users to choose. 

\section{Implementation of the gPROMS-MPC tool}\label{sec:implementation_gMPC}
The implementation of the gPROMS-MPC tool is based on the gPROMS Foreign Process Interface, which is a C/C++ interface to allow developers' custom code to interact dynamically with the gPROMS model at runtime. Although many optimization algorithms can solve MPC problems, the online successive linearization strategy used in the gPROMS-MPC tool, clearly encourages us to implement our previous construction-free \textit{CDAL} algorithm \cite{wu2021simple} and a parametric active-set \textit{qpOASES} v3.2 algorithm \cite{ferreau2014qpoases} from efficiency perspective. The \textit{CDAL} and \textit{qpOASES} solve the sparse (\ref{sparseQP}) and condensed (\ref{condensedQP}) QP formulations, respectively.

The \textit{CDAL} is based on the coordinate descent (CD) and augmented Lagrangian (AL) methods. The outer loop involves the accelerated AL iteration, and the inner loop (solving the AL subproblem) uses the CD method, exploiting the structure of the MPC problem. And an efficient CD-AL coupling scheme and preconditioner are proposed in the implementation of the \textit{CDAL} to speed up computation \cite{wu2021simple}. Its most notable feature is that the \textit{CDAL} directly solves the MPC formulation (\ref{MPC}) without resorting to an explicit MPC-to-QP construction, which is called \emph{construction-free}. Clearly, the \emph{construction-free} feature is suitable for the online successive linearization strategy used in the gPROMS-MPC tool to avoid the online construction cost. Especially in some online successive linearization based MPC problems, its online construction cost is comparable to the online solving itself thanks to warm-starting, gradually changing set-points and slowly updating dynamics. In addition, our \textit{CDAL} is also \emph{matrix-free} (avoids multiplications and factorizations of matrices) and \emph{library-free} (without any library dependency), its 90-lines C-code implementation makes it competent in the work of embedded MPC code generation. The detailed description and implementation of the \textit{CDAL} algorithm are presented in \cite{wu2021simple}.

Although gPROMS provides minimal state-space realization, its derived linear state-space model still involves large state dimensions for some plants, especially when it involves spatial-time equations. In these cases, solving the condensed QP formulation associated with MPC is a better choice. The condensed QP formulation (\ref{condensedQP}) is solved by calling the \textit{qpOASES} package, which is an open-source C++ implementation of the online parametric active set strategy \cite{ferreau2014qpoases}. The \textit{qpOASES} not only supports warm-starting strategy and also supports solving QPs with time-varying Hessian matrices, that makes it suitable for the online successive linearization-based MPC problems. In addition to integrating the called \textit{qpOASES} package for efficiently online solving QPs, the gPROMS-MPC tool implements efficiently the online condensed MPC-to-QP construction. Herein the main condensed equation listed in (\ref{condensedEqn}),
\begin{equation}\label{condensedEqn}
\left[\begin{array}{c} 
\delta Y_1\\
\delta Y_2\\
\vdots\\
\delta Y_{T}
\end{array}\right]
= G\left[\begin{array}{c} 
\hat{X}_1\\
\hat{X}_2\\
\vdots\\
\hat{X}_{T}
\end{array}\right]
=GM\left[\begin{array}{c} 
\hat{U}_0\\
\hat{U}_1\\
\vdots\\
\hat{U}_{T-1}
\end{array}\right]
+Gm
\end{equation}
where $M = \left[\begin{array}{cccc}
\hat{B}     &  0  & \cdots & 0\\
\hat{A}\hat{B}     & B & \cdots & 0\\
\vdots & \vdots & \ddots & \vdots\\
\hat{A}^{T-1}\hat{B} & \hat{A}^{T-2}\hat{B} & \cdots & \hat{B}
\end{array}
\right]$,\\
$m =\left[\begin{array}{c}
     e\\
     \hat{A}e+e\\
     \vdots \\
     \hat{A}^{T-1}e+\hat{A}^{T-2}e+\cdots+e
\end{array}\right]$,\\
$G = \left[\begin{array}{cccc}
\hat{C}     &  0  & \cdots & 0\\
0     & \hat{C} & \cdots & 0\\
\vdots & \vdots & \ddots & \vdots\\
0 & 0 & \cdots & \hat{C}
\end{array}
\right]$.\\
Then, the dense hessian matrix $H_c$ and gradient vector $h_c$ of the condensed QP formulation (\ref{condensedQP}) are
\begin{equation*}
\begin{aligned}
H_c &= (GM)^{\prime}\left[\begin{array}{cccc}
W^y     &  0  & \cdots & 0\\
0     & W^y & \cdots & 0\\
\vdots & \vdots & \ddots & \vdots\\
0 & 0 & \cdots & W^y
\end{array}
\right](GM) \\
&+ \left[\begin{array}{cccc}
W^{\Delta u}    &  0  & \cdots & 0\\
0     & W^{\Delta u} & \cdots & 0\\
\vdots & \vdots & \ddots & \vdots\\
0 & 0 & \cdots & W^{\Delta u}
\end{array}
\right]
\end{aligned}
\end{equation*}
\begin{equation*}
\begin{aligned}
h_c&=(GM)^{\prime}\left[\begin{array}{cccc}
W^y     &  0  & \cdots & 0\\
0     & W^y & \cdots & 0\\
\vdots & \vdots & \ddots & \vdots\\
0 & 0 & \cdots & W^y
\end{array}
\right]\left(Gm-\left[\begin{array}{c}
     \delta r\\
     \delta r\\
     \vdots\\
     \delta r
\end{array}\right]
\right)
\end{aligned}
\end{equation*}
To exploit the above structures, the gPROMS-MPC tool implements an efficient condensing procedure, which shares the same idea with the Andersson et al. work \cite{andersson2013condensing} via the use of OpenBLAS v0.3.20 \cite{xianyi2016openblas}.

\section{Application example of Flash-Separation}\label{sec:application_example}
\begin{figure}
\centering
        \hspace*{-1em}\includegraphics[width=1\columnwidth]{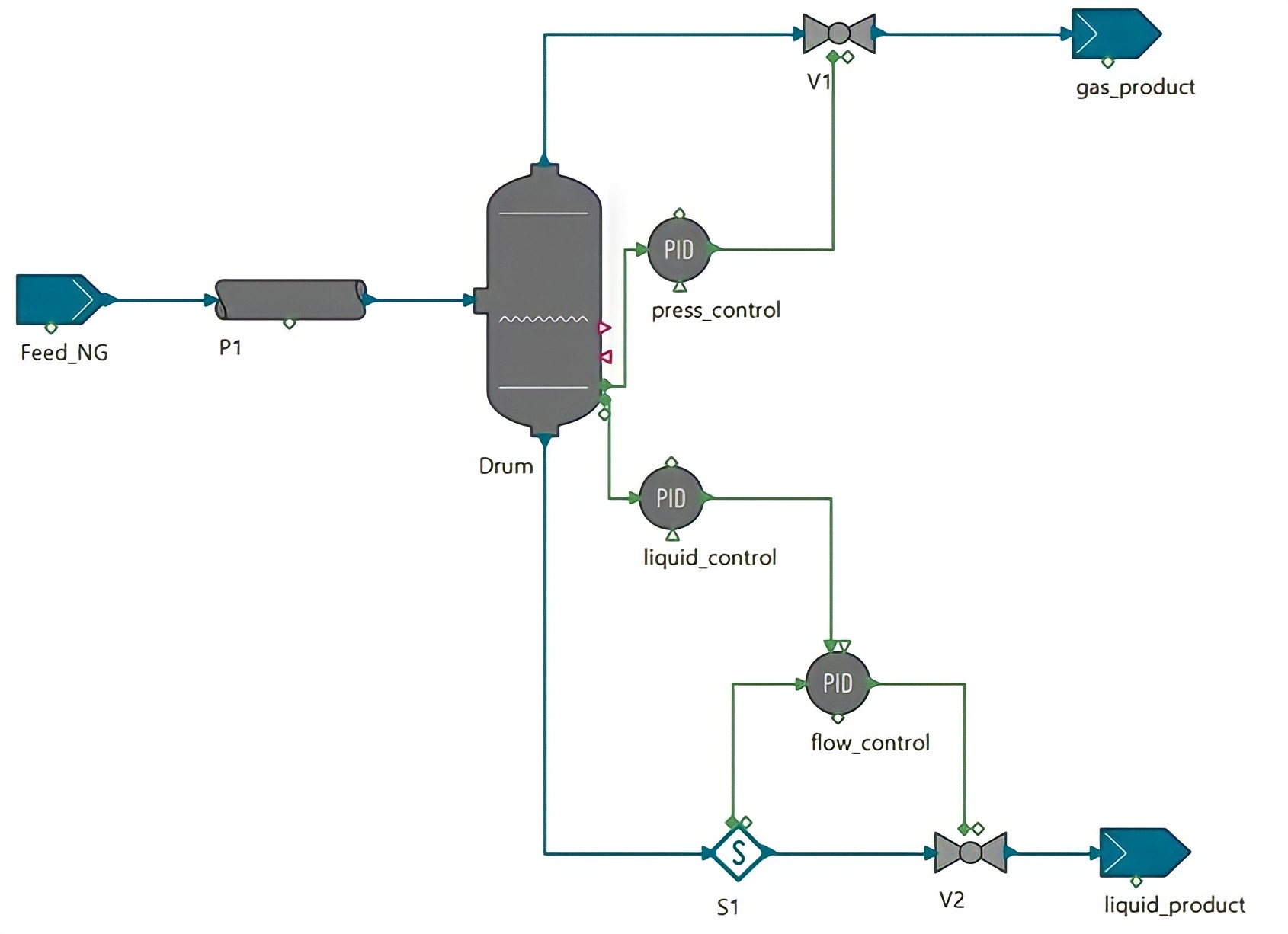} 
        \caption{Flash separation flowsheet under the continuous-time PID control scheme}
        \label{fig:flowsheet_pid} 
\end{figure}

The developed gPROMS-MPC tool worked well on some MPC benchmark examples in the literature such as the ill-conditioned AFTI-16 \cite{bemporad1997nonlinear} and nonlinear CSTR \cite{seborg2010process}, which demonstrates the effectiveness of its functionality. To further validate the correctness of the gPROMS-MPC tool when applying to the commercial built-in gPROMS Model Library (gML), this section presents the MPC controller design for a flash separation of a mixture of eight components. To investigate performances of different control schemes, three control schemes are used for comparison, namely multi-PID scheme, successive-lineariztion-based MPC (SL-MPC) scheme, and linear MPC scheme. The multi-PID scheme is built as shown in Fig. \ref{fig:flowsheet_pid}, which involves one source model (Feed\_NG), one pipeline model (P1), one flash drum model (Drum),  two value models (V1 and V2), two sink models (gas\_product and liquid\_product), one stream analyzer model (S1) that only pass the mass flowrate and three continuous-time PID controller models (liquid\_control and flow\_control are connected as the cascade PID). The parameters of the flowsheet with PID scheme is listed in Table \ref{tab1}. 

In the SL-MPC and linear MPC schemes, the MPC controller is used to replace three PID models and the stream analyzer model, and other model parameters remain the same. The MPC controller directly manipulates the stem position of two valve simultaneously to regulate the pressure and liquid level fraction of the Drum, namely a two-input-two-output plant. As for the MPC settings, the prediction horizon is 10, the sampling time is $0.02$ s, the constraints come from the physical constraints of the two values V1 and V2, that is box-constraints $[0,1]$ The cost weights of the increased input and the output are $W^{\Delta u}=diag([1,1])$ and $W^y=diag([100,100])$, respectively. The difference of the SL-MPC and linear MPC scheme is that, the former updates the linear state-space model at each sampling time via linearization along the trajectory, and the latter only utilizes one linear state-space model that is linearized at initial point. As we mentioned in Section \ref{sec:intro}, even though many manufacturing processes are inherently nonlinear, the performance of linear MPC scheme often satisfies requirements in the vast majority of process industry applications. Thus, obtaining a good linear state-space model, with good close-loop performance in operating ranges, is the key part, which is also most time-consuming and labor-intensive step in whole MPC design workflow. That is what the gPROMS-MPC tool does, being an MPC design platform based on gPROMS' powerful modeling capability.

\begin{table*}[!htbp]
\caption{The settings of flash separation with PID scheme}
\centering
\begin{tabular}{cc}
\toprule
model name  & model parameters\\
\midrule
{Feed\_NG} & components name and mass fraction:\\
& [METHANE=0.5202], [ETHANE=0.1504], [PROPANE=0.1180]\\
& [I-BUTANE=0.0167],[N-BUTANE=0.0462], [I-PENTANE=0.0098]\\
& [N-PENTANE, 0.0090],[N-HEXANE, 0.0006]\\
& pressure = 70 bar, Vapour mass fraction = 0.95\\
\midrule
{P1}& Inner diameter = 0.152606 m, length = 10 m\\
& Friction facotr correlation: Fully turbulent\\
\midrule
{Drum} & Shape: Drum, Orientation: Vertical, Diameter = 0.5 m, Volume = 3 $m^3$\\
& Include effect of hydrostatic pressure\\
\midrule
{V1} & Phase: Vapour, Flow relation: Fisher universal gas sizing equation\\
& Flow coefficient = 200000 $scf h^{-1} psi^{-1}$, Recobery factor = 34.8\\
& Inherent characteristic: Linear, Leakage fraction = 0\\
\midrule
{V2} & Phase: Liquid, Flow relation: Fisher equation\\
& Flow coefficient = 100 $scf h^{-1} psi^{-1}$\\
& Inherent characteristic: Linear, Leakage fraction = 0\\
\midrule
{pressure\_control}& Controller action: Direct, Controller gain = 1000, Integral time constant = 5\\
& Process variable: Pressure, Manipulated variable: Stem position\\
& initial setpoint: 69 \\\midrule
liquid\_control & Controller action: Direct, Controller gain = 10, Integral time constant = 1\\
& Process variable: Liquid level fraction, Manipulated variable: mass flowrate \\
& initial setpoint: 0.5 \\
\midrule
flow\_control & Controller action: Reverse, Controller gain = 100, Integral time constant = 10\\
& Process variable: Mass flowrate, Manipulated variable: Stem position\\
& setpoint: from liquid\_control\\
\bottomrule
\end{tabular}
\label{tab1}
\end{table*}

\begin{figure*}[htbp]
\centering
\subfigure[Liquid level fraction v.s. time]{
\begin{minipage}{7cm}\label{fig:3:a}
\centering
\includegraphics[scale=0.6]{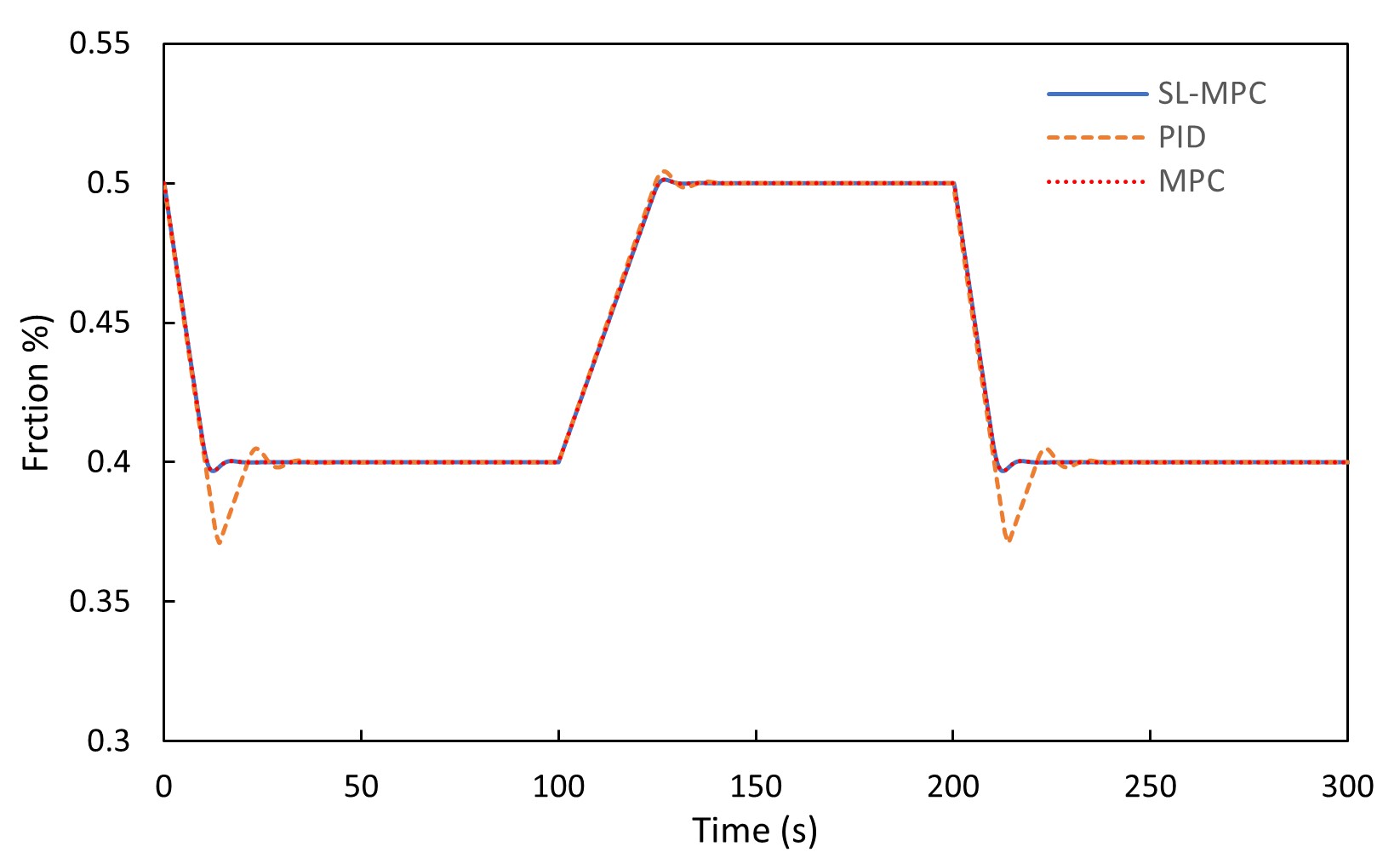}
\end{minipage}
}
\subfigure[Pressure v.s. time]{
\begin{minipage}{7cm}\label{fig:3:b}
\centering
\includegraphics[scale=0.6]{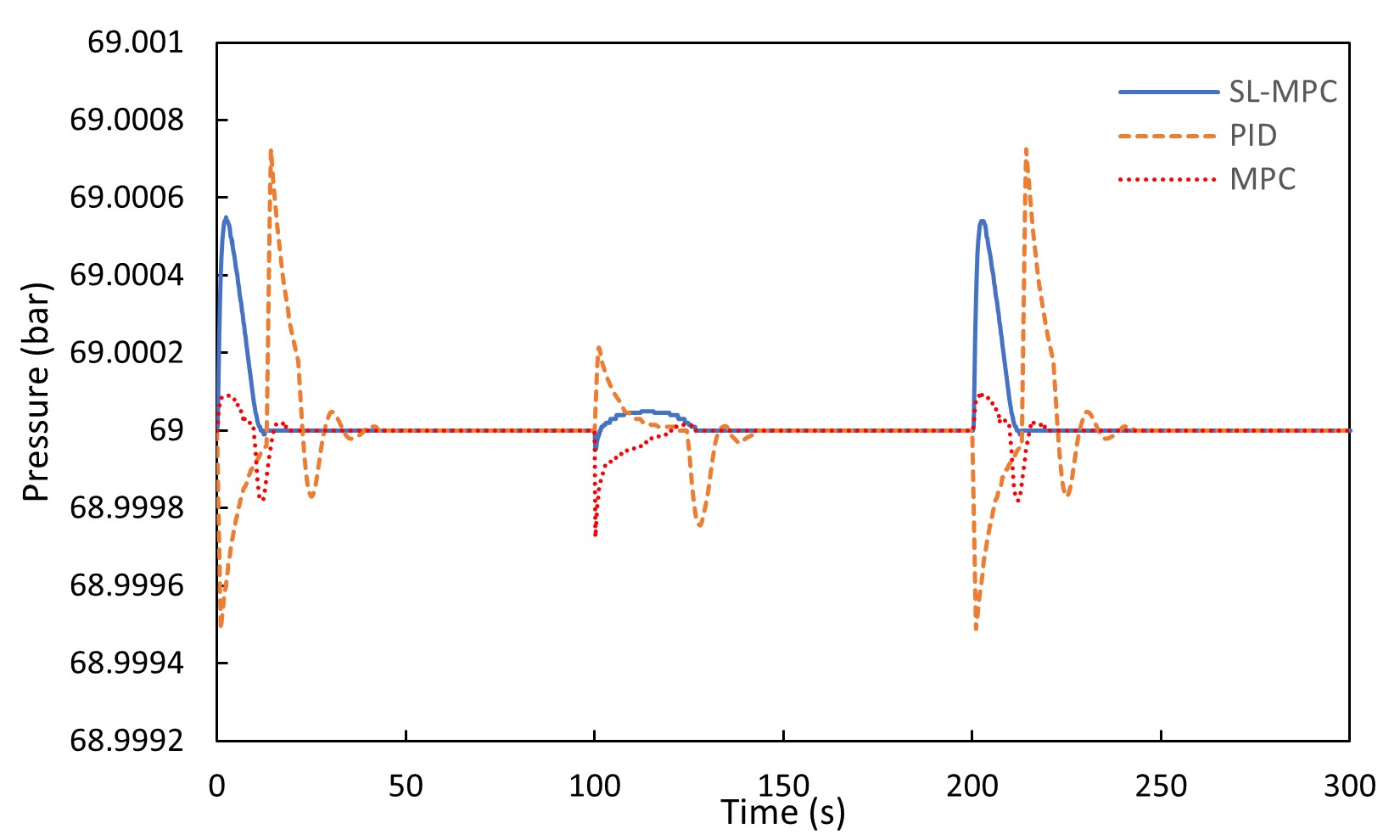}
\end{minipage}
}
\subfigure[Liquid mass flowrate v.s. time]{
\begin{minipage}{7cm}\label{fig:3:c}
\centering
\includegraphics[scale=0.6]{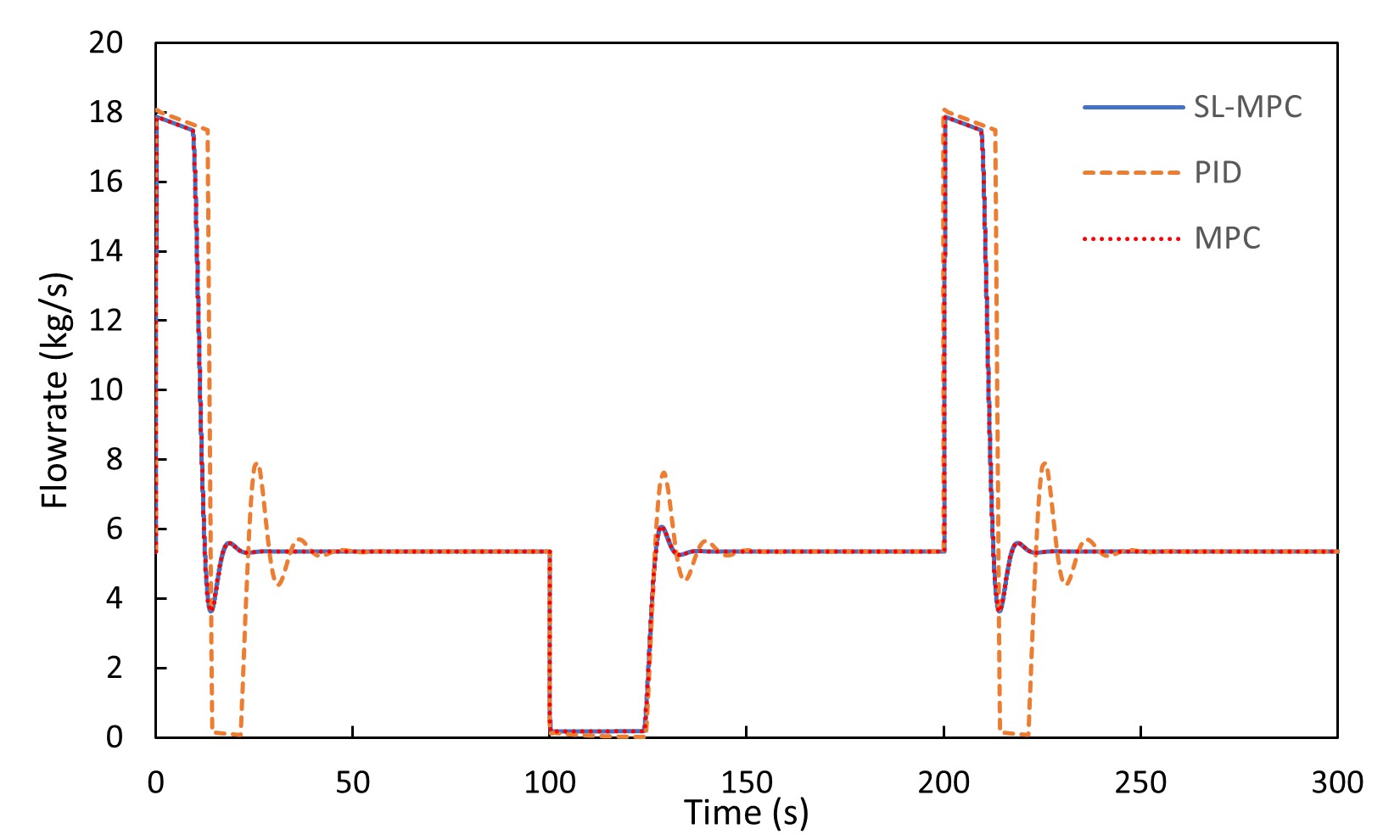}
\end{minipage}
}
\subfigure[Vapour mass flowrate v.s. time]{
\begin{minipage}{7cm}\label{fig:3:d}
\centering
\includegraphics[scale=0.6]{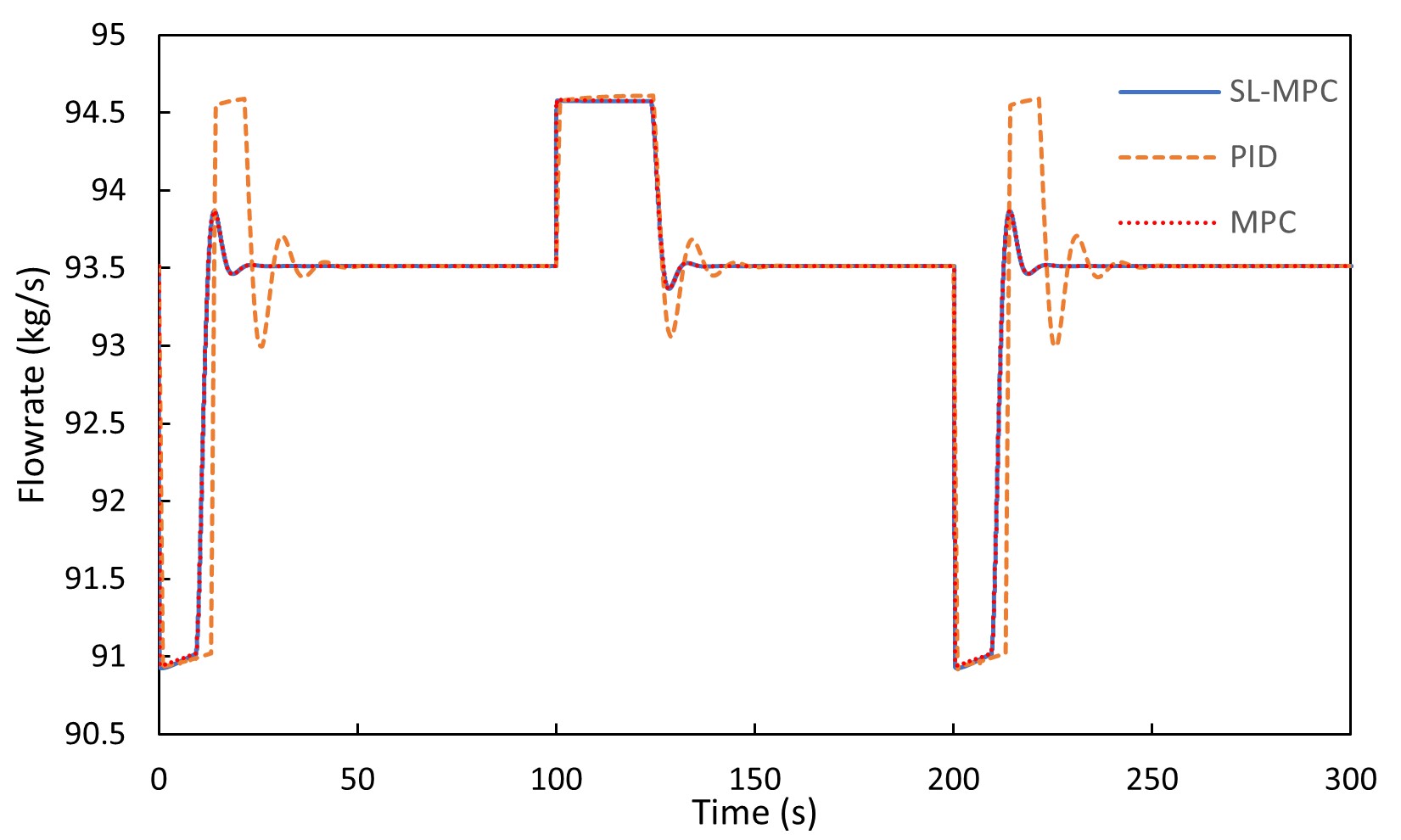}
\end{minipage}
}
\caption{Comparison results of PID, SL-MPC and MPC scheme in the flash-separation example}
\label{fig:3}
\end{figure*}

The simulation scenario is to track the desired setpoints of liquid level and pressure fraction from the initial steady-state condition in which pressure $= 69$ bar and liquid level fraction $= 0.5$. The simulation time is from 0 to 300 second, and the desired setpoints of liquid level fraction are 0.4, 0.5 and 0.4 at every 100 second, and the desired setpoint of pressure keeps at 69 bar all the time. The comparison simulation results are shown in Fig. \ref{fig:3:a}, \ref{fig:3:b}, \ref{fig:3:c} and \ref{fig:3:d}, which are liquid level fraction, pressure, liquid mass flowrate and vapour mass flowrate, respectively. The results show that the closed-loop control performance of the SL-MPC and MPC scheme are better than the PID scheme, having less integral overshoot and oscillation, and the SL-MPC and MPC scheme have almost the same tracking performance, which can be explained that the nonlinearity of the flash-separation example is not strong in operating ranges. It shows that the linearized model at initial equilibrium point can be used in embedded MPC code generation.

\section{Conclusion and future works}\label{sec:conclusion}
This paper develops the gPROMS-MPC tool, which not only interacts directly with the gPROMS first-principle-based model for closed-loop simulations but also utilizes its mathematical information to derive the linearized state-space model for MPC design. The gPROMS-MPC tool adopts the online successive linearization-based MPC to handle general nonlinear systems and also allows users to choose when to linearize such as at some specific points to obtain one or multiple linear models for later embedded MPC design. The gPROMS-MPC tool implements our previous construction-free \textit{CDAL} and the online parametric active-set \textit{qpOASES} algorithm. Our \textit{CDAL} is also matrix-free and library-free, which provides benefits in embedded C-code generation. 

The derived linear model from the gPROMS is state-space formulation (input-state-output), which would suffer high state-dimension issues and require a state estimation algorithm in embedded MPC deployment. They could be addressed to some extent by using the model reduction technique to reduce state dimension, but we recommend transforming it into an equivalent input-output ARX model as illustrated in our work \cite{wu2022equivalence}, to avoid designing state estimation algorithm. And the resulting ARX-based MPC problem can be solved by our extended \textit{CDAL-ARX} algorithm \cite{wu2022construction}. Another future work will focus on extending the gPROMS-MPC tool to solve economic MPC problems, and further apply it to more practical industrial applications.
\bibliographystyle{unsrt}
\bibliography{ref} 
\end{document}